\documentclass[aps,prb,preprint,showpacs,groupedaddress]{revtex4}
\usepackage{graphicx}  
\usepackage{dcolumn}   
\usepackage{bm}        
\usepackage{amssymb}   

\begin{document}

\title{Anharmonicity of vibrational modes in fullerenes}
\author{Hengjia Wang}
\author{Murray S. Daw}
\affiliation{Clemson University, Clemson, SC, USA 29634-0978} 

\begin{abstract}
 We report a computational study of the anharmonicity of the vibrational modes of various individual fullerenes using the ``moments method'' [Y. Gao and M. Daw, \textbf{Modelling Simul. Mater. Sci. Eng. 23} 045002 (2015)] with a Tersoff-style potential for carbon. We find that the frequencies of all vibrational modes drop systematically with temperature and the sizes of the individual fullerenes do not affect strongly the anharmonicity of the modes. Comparison is made with available experiments.

\end{abstract}

\pacs{61.48.-c, 65.80.-g, 02.70.Uu, 31.15.bu}

\maketitle

\section{Introduction}

Vibrational modes of fullerenes are accessible experimentally by Raman spectroscopy~\cite{Raman1, Raman2}, IR scattering~\cite{IR1} and inelastic neutron scattering~\cite{ins1}. The complete assignment of the vibrational modes of $C_{60}$, the most popular fullerene, has been made experimentally by these three approaches and also theoretically using, for example, the DFT method.~\cite{ins1, DFT2}

By examining the temperature dependence of the spectra, it is possible to explore the anharmonicity of the active modes. Raman spectra of $C_{60}$ have revealed that the frequency of all the peaks drops with temperature.~\cite{Texp} In this work, we apply the ``moments method''\cite{momentstheory, jazzpaper, Jazznote} to investigate the temperature-dependence of the frequency of all of the modes of individual fullerenes. We also study how the size of the individual fullerene affects the anharmonicity of the modes. We find that the value of the anharmonicity (measured by $\omega'(T)/\omega(0)$) of all the modes for the fullerenes is about $-3 * 10^{-5} K^{-1}$, largely independent of the mode or the size of the fullerene. 

We begin with a brief introduction to the ``moments method'', and then show our results for the anharmonicity of the vibrational modes in fullerenes, ending with a summation.

\section{Method}

The moments method is an approximation based on low-order moments of
the Liouvillian operator~\cite{momentstheory}, which is the
time-evolution operator of phase-space functions for a \emph{classical} dynamical system. Beginning
with the harmonic force constant matrix for the particular cell, the normal modes are found,
indexed by $b$. The calculation involves ensemble averaging of products of normal mode amplitudes $A_{b}$ and accelerations $\ddot{A}_{b}$, which are obtained by projecting the atomic displacements and forces onto the normal modes. Using the harmonic modes as a basis is justified by the weakly anharmonic character of this system. The lowest, non-trivial moment of the power spectrum of the displacement-displacement autocorrelation 
\begin{equation}
\mu_2(b) = -\frac{\langle A_{b} \ddot{A}_{b} \rangle}{\langle A_{b}^2 \rangle - \langle A_{b} \rangle ^ 2}
\label{eq:mu2}
\end{equation}
(where the angle brackets indicate ensemble averages) gives a simple measure of the temperature-dependent dynamics of the system.

The previous expression includes the possibility that $\langle A_{b} \rangle$ is non-zero, which was not included in some of our previous papers because the systems considered previously had sufficiently high symmetry that the average displacement vanished. However, in the present case, the average displacement of the RBM deviates significantly from zero as the temperature increases, and so we have extended the expressions derived in previous work to include non-zero first moments.

The quasi-harmonic (temperature-dependent) frequency $\omega$ is given by 
\begin{equation}
\omega(b) = \sqrt{\mu_2(b)} 
\label{eq:omega}
\end{equation}
The moment is calculated by standard Monte Carlo integration. This method was used recently to study the anharmonic renormalization of flexural modes in graphene~\cite{Wang2016} and the anharmonicity of the RBM in SWCNTs~\cite{Wang2017}. 

In this study, the interatomic interaction is described by a Tersoff potential tailored somewhat for graphene~\cite{TersoffPRL,origTersoff,modTersoff}. The normal modes are identified from the eigenvectors of the harmonic force constant matrix. The number of Monte Carlo steps is $10^6$ for the calculation on each individual fullerene. In some parts of the analysis, we focus on the RBM, but all modes have been included in the calculation. For all the calculations, the sides of the cubic simulation box are large enough so that the interaction between individual fullerenes could be neglected. The fullerenes involved in the calculations are $C_{20}$, $C_{36}$, $C_{60}$, $C_{70}$, $C_{84}$, $C_{100}$, $C_{180}$ and $C_{240}$. For those fullerenes that have allotropes, we chose to focus on the most symmetrical structures. For the ellipsoid-shaped fullerenes, the mode in which the fullerene contracts or expands uniformly is treated as the RBM, although its symmetry is not $A_g(1)$. For example, for the $C_{70}$ fullerene we call the $A_1^{'}(2)$ mode the RBM.

\section{Results}

\begin{figure}
\includegraphics[width=\columnwidth]{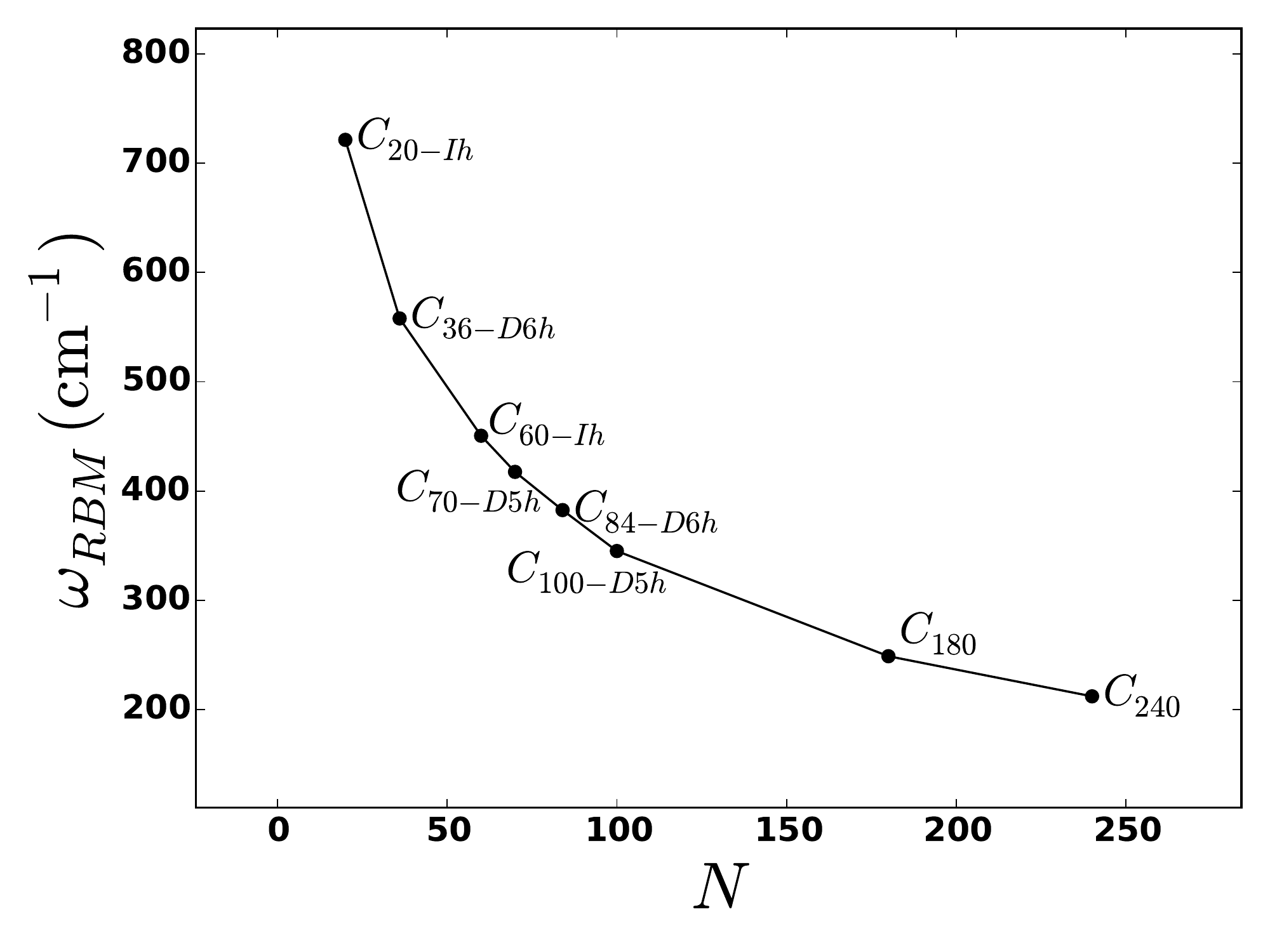}
\caption{\label{fig:inverse} The dependence of the harmonic frequency of the RBMs for various individual fullerenes on the number of atoms $N$ in the fullerene. }
\end{figure}

\begin{figure}
\includegraphics[width=\columnwidth]{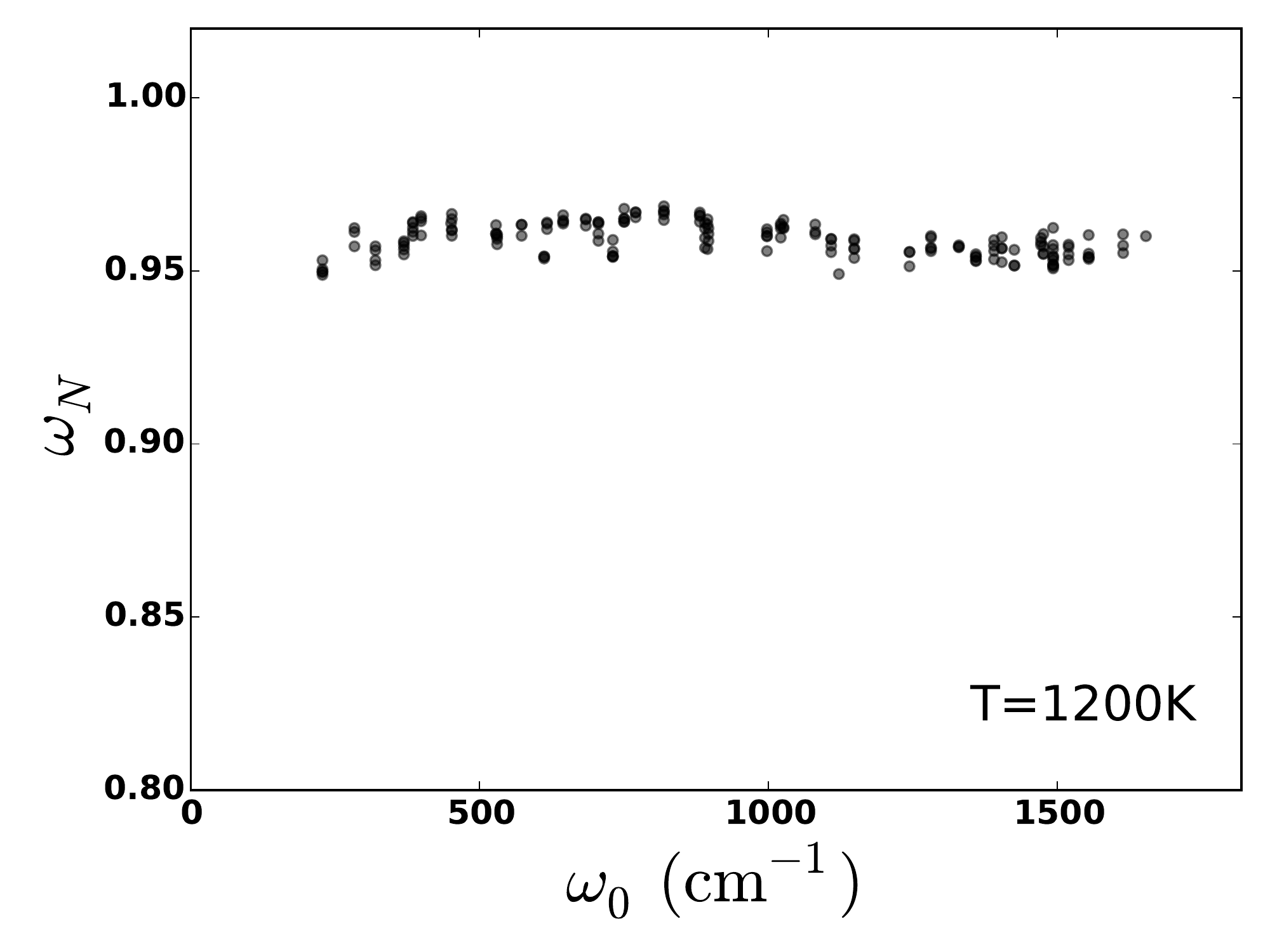}
\caption{\label{fig:scatter} Scatterplot of the ratio $\omega_N \equiv \omega/\omega_0$ vs. $\omega_0$ for all modes of the $C_{60}$ fullerene at $T=1200K$. For a perfectly harmonic system, all points would be at the top (at a value of 1); due to anharmonicity, all of the modes of this system drop to lower frequency with increasing temperature. The more anharmonic modes exhibit lower values of $\omega_N$.}
\end{figure}

\begin{figure}
\includegraphics[width=\columnwidth]{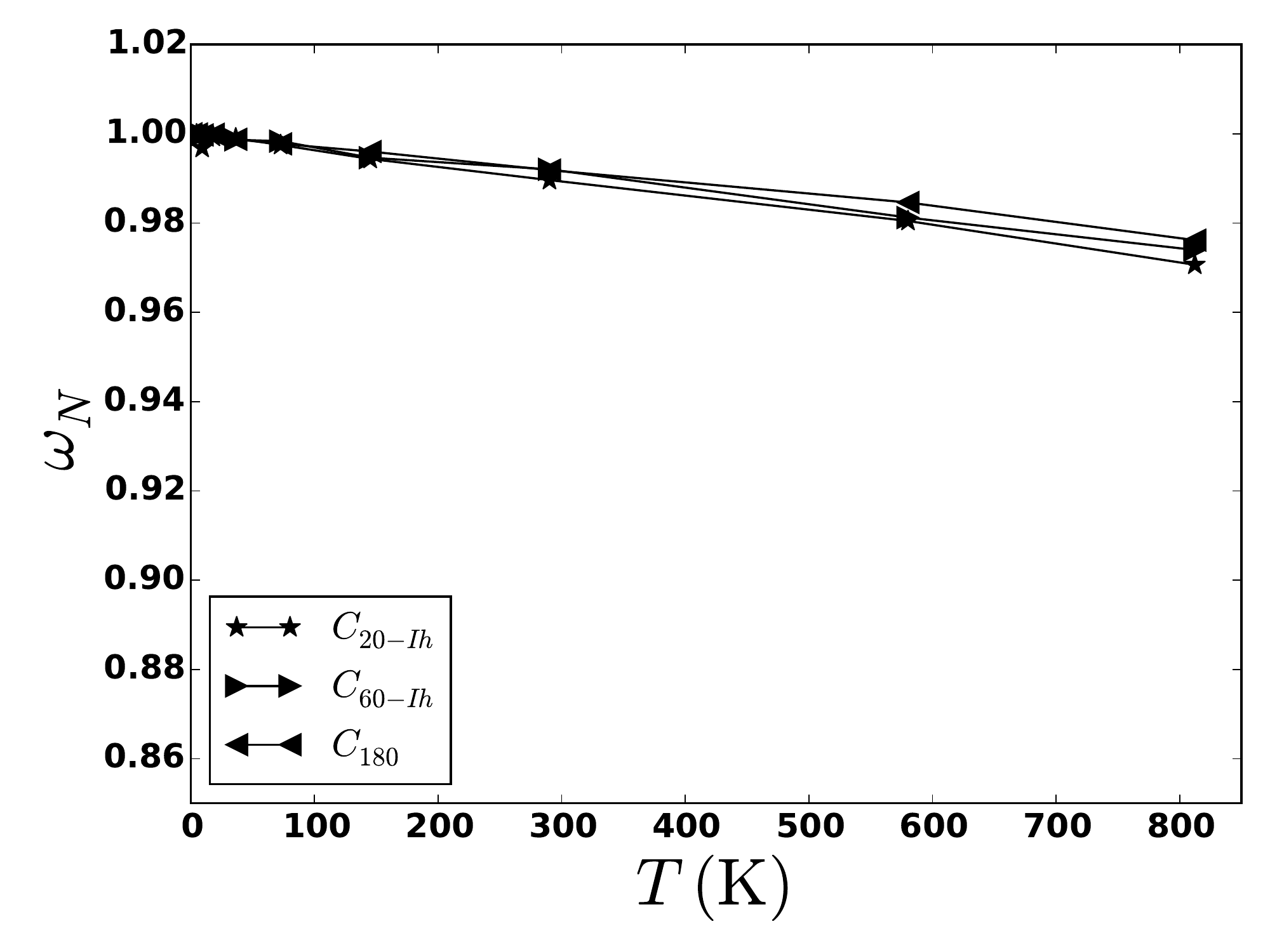}
\caption{\label{fig:curves} Plot of the ratio $\omega_N \equiv \omega/\omega_0$ vs. temperature $T$ for the RBMs of three fullerenes.}
\end{figure}

We begin by presenting the results of the harmonic frequencies of the RBMs for the individual fullerenes. Fig.~\ref{fig:inverse} shows that the harmonic frequency of the RBM has an inverse relationship with the size of the fullerene (measured by the number of the atoms in the molecule.) This behavior is similar to that seen for the RBM of SWCNTs.~\cite{Maultzsch2005} Our harmonic results resemble and serve as a calibration against earlier experimental results and other calculational results that use different methods or different empirical potentials.~\cite{harmonic1, harmonic2, harmonic3} However, the main focus of this paper is on the anharmonicity of the modes, as we will discuss next.

In terms of the anharmonicity of the vibrational modes, we find that the frequencies of all the modes for all the fullerenes drop roughly linearly with temperature, and in proportion to frequency as well. We therefore define as a measure of the anharmonicity \[ p = \frac{1}{\omega_0}\frac{d \omega}{dT} \] so that a perfectly harmonic system would have $p=0$. Our calculated value of $p$ for all the modes falls between $-2 * 10^{-5} K^{-1}$ and $-4 * 10^{-5} K^{-1}$ for all the fullerenes we studied here. As an example, Fig.~\ref{fig:scatter} is a scatterplot of the ratio $\omega_N \equiv \omega/\omega_0$ vs. $\omega_0$ for all modes of the $C_{60}$ fullerene at $T=1200K$. All modes show similar temperature dependence. In comparison, slopes between $-0.8 * 10^{-5} K^{-1}$ and $-2 * 10^{-5} K^{-1}$ are observed experimentally~\cite{Texp} for the shifts of the peaks in Raman spectra of solid $C_{60}$. The comparison between our calculations and the experiment is reasonably good; the differences could be attributed to many aspects, such as individual $C_{60}$ vs. solid $C_{60}$ and also the quality of the empirical potentials, which are usually not determined by reference to anharmonic properties. 

Inspired by the high anharmonicity of the RBM in SWCNTs ~\cite{Wang2017}, we pick out the RBMs in fullerenes and study their anharmonicity. Fig.~\ref{fig:curves} shows the temperature dependence of the frequency of RBMs of three fullerenes ($C_{20-Ih}$, $C_{60-Ih}$, and $C_{180}$) that were typical of the rest in this regard. The results for the other fullerenes we considered ($C_{36-D6h}$, $C_{70-D5h}$, $C_{84-D6h}$, $C_{100-D5h}$, and $C_{240}$) are very similar and were not included in the plot because they overlap significantly with the three shown. The value of $p$ does not appear to vary much with the size of the fullerene. (In contrast, we have observed in SWCNTs that the value of $p$ of the RBMs increases in magnitude with diameter of the tube.~\cite{Wang2017})

\section{Conclusions}

We have presented the results of a study of anharmonicity of vibrational modes of fullerenes obtained by the ``moments method'', which is based on Monte Carlo averages of products among displacements and forces. The forces and energies required for the MC calculation were obtained from a semi-empirical Tersoff potential, somewhat optimized for graphene-like structures. 

Our results for the harmonic part resemble earlier works, showing that the harmonic frequency of the RBM of the fullerene is inversely related to the size of the fullerene. With regards to anharmonicity, generally all modes, including the RBMs, shift down in frequency with increasing temperature at roughly similar speeds for all fullerenes we have used. The sizes of the individual fullerenes do not affect strongly the anharmonicity of the modes. The results provide a clearer picture of the anharmonicity of the vibrational modes of fullerenes. 
  
\section{Acknowledgement}
Our research was supported by the US Department of Energy, Office of Basic Energy Science, Division of Materials Sciences and Engineering under Grant No. DE-SC0008487.

\end{document}